\PassOptionsToPackage{prologue,dvipsnames}{xcolor}
\documentclass[sigplan,10pt,nonacm,screen]{acmart}

\usepackage[english]{babel}
\usepackage{blindtext}
\usepackage{epsfig,endnotes}
\usepackage{graphicx}
\usepackage{xspace}
\usepackage[normalem]{ulem}
\usepackage{multirow}
\usepackage{array}
\usepackage{enumitem}

\setlist{nosep}
\usepackage{pifont}
\usepackage{subcaption}
\usepackage{algorithm}
\usepackage[noend]{algpseudocode}
\usepackage{listings}
\usepackage{amsmath} 
\usepackage{hyperref}
\hypersetup{colorlinks={true},linkcolor={violet},citecolor=blue}
\usepackage{cleveref}
\usepackage[font=small]{caption}
\usepackage[font=small]{subcaption}
\usepackage[dvipsnames]{xcolor}
\usepackage{pythonhighlight}
\crefname{section}{}{\S\S}
\crefdefaultlabelformat{\S#2#1#3}
\Crefformat{figure}{#2{}Figure #1{}#3}

\newcommand{\sys}{{{DSV}}\xspace}
\newcommand{\topk}{{{top-$k$}}\xspace}

\renewcommand\footnotetextcopyrightpermission[1]{} %
\setcopyright{none}

\acmDOI{}

\acmISBN{}

\acmConference[]{}
\acmYear{}
\copyrightyear{}

\acmPrice{}

\begin{document}

\title{\sys: Exploiting Dynamic Sparsity to Accelerate Large-Scale Video DiT Training}

\author{
    Xin Tan$^1$,
    Yuetao Chen$^1$,
    Yimin Jiang$^3$,
    Xing Chen$^2$,
    Kun Yan$^2$,
    Nan Duan$^2$,\\
    Yibo Zhu$^2$,
    Daxin Jiang$^2$,
    Hong Xu$^1$
}
\affiliation{
    \institution{$^1$The Chinese University of Hong Kong, $^2$StepFun, $^3$Unaffiliated}
    \city{}
    \country{}
}

\renewcommand{\shortauthors}{}

\begin{abstract}
Diffusion Transformers (DiTs) have shown remarkable performance in generating high-quality videos. However, the quadratic complexity of 3D full attention remains a bottleneck in scaling DiT training, especially with high-definition, lengthy videos, where it can consume up to 95\% of processing time and demand specialized context parallelism.

This paper introduces \sys to accelerate video DiT training by leveraging the dynamic attention sparsity we empirically observe. \sys uses a two-stage algorithm to capture the dynamic sparsity patterns via low-rank based approximation of the original query and key. It employs custom kernels to efficiently identify critical key-value pairs and compute the sparse attention. 
To accommodate the new sparsity dimension, \sys adopts a hybrid sparsity-aware context parallelism that re-balances the skewed workload across attention heads and blocks due to sparsity heterogeneity. \sys achieves up to 3.02$\times$ higher training throughput, scaling to 128 GPUs and 520k token lengths, without quality loss.

\end{abstract}

\maketitle
\pagestyle{plain}

\section{Introduction}

Text-to-video generation has experienced significant breakthroughs recently and received widespread interest~\cite{esser2024stablediff3, chen2023pixart, peebles2023scalableDiT, ma2024latte, klingvideo, HunyuanVideo, bao2024vidu, polyak2024movie}. Behind these advancements is the Diffusion Transformers (DiTs)~\cite{peebles2023scalableDiT,ma2024latte,Openai-sora}, which have emerged as state-of-the-art architectures within diffusion-based generative paradigms~\cite{ho2020denoisingddpm,lipman2022flowmatching}. These paradigms learn data distributions by progressively adding noise to input samples and training a model to reverse this process. During inference, the model starts with random noise and iteratively denoises it to generate high-quality videos.

In contrast to large language models (LLMs) \cite{openai2023gpt4,llama-3-1}, which use transformer also but scale to hundreds of billions of parameters, DiTs are smaller, typically under tens of billions of parameters \cite{OpenSora,yang2024cogvideox,polyak2024movie,HunyuanVideo}. While their smaller scale reduces some challenges in training, video DiTs still struggle in processing long, high-resolution video inputs---a growing concern fueled by expanding datasets and applications like film post-production and multi-camera event capture \cite{Openai-sora,polyak2024movie}. For example, even with compressed latent video representations from variational autoencoders (VAEs) \cite{stabilityai_stable_diffusion}, token length for high-definition or long sequences can easily reach hundreds of thousands.

The primary bottleneck lies in the attention module~\cite{waswani2017attention}, which has quadratic time complexity with respect to input length and can consume over 80\% of training time (Figure~\ref{fig:attention_bottleneck}). This issue worsens when the latent token length exceeds hundreds of thousands, as a single GPU cannot process the entire sequence in memory. 
{Context parallelism}~\cite{liu2023ring,fang2024usp,jacobs2023deepspeedulysses,gu2024loongtrain} addresses this by distributing the input across GPUs for parallel processing, but it introduces  additional challenges with inter-device communication.

Fortunately, our empirical observations reveal that attention computations in video DiTs exhibit notable sparsity that may mitigate the attention bottleneck~\cite{zhang2023h2o,xiao2023streamllm}. Attention scores tend to follow a power-law distribution, with a few key-value (KV) pairs dominating the total score. However, unlike some predictable attention patterns observed in LLMs (e.g. attention sinks or window pattern~\cite{mistral7b,xiao2023streamllm}), we observe that attention sparsity in video DiTs is \textit{dynamic}: 
The critical KV positions in video DiTs do not have clear locality or other patterns.
The sparsity level and pattern varies not only across blocks but also across attention heads within the same block. 
Finally, sparsity also intensifies progressively during training. 
These observations, reported for the first time to our knowledge, show that methods assuming fixed sparsity patterns are ineffective for DiT training, and a \textit{dynamic} approach is needed to reap attention sparsity effectively.

A natural solution to leverage dynamic sparsity is to compute the attention scores and find critical KV at each attention operation. 
This entails two problems. 
First, large-scale training relies on I/O-aware fused kernels for maximum efficiency~\cite{dao2022flashattention1,dao2023flashattention2}. 
Extracting the full attention matrix disrupts these optimizations and causes significant slowdown. 
Second, even if critical KV pairs are identified, most attention computation (e.g., softmax score) would already be completed, which eliminates most performance benefits. 
Additionally, current context parallelism~\cite{jacobs2023deepspeedulysses,liu2023ring} cannot effectively capture the heterogeneous sparsity across attention heads, leading to suboptimal performance.

Based on these insights, we present \sys, a framework that accelerates the video DiT training by exploiting dynamic sparsity patterns in attention while maintaining model quality. The core idea of \sys is to approximate attention scores using distinct predictors for each attention head. This facilitates the identification of critical KV pairs without tearing up the fused attention kernels.
Attention can then be computed only on these critical pairs, and context parallelism can be adjusted accordingly to only exchange information for them. Specifically, \sys comprises three key components:

First, \sys employs a two-stage DiT training approach to leverage dynamic sparsity. In the first stage, it trains low-rank predictors to approximate $QK^T$ for each attention head independently from the DiT training. Once the predictors are trained, \sys transitions to the second stage: It dynamically assesses the cost-benefit trade-off to activate sparse attention on individual blocks based on their profiled sparsity levels. 
It then uses the predictors to estimate the critical KV pairs.

Second, to efficiently carry out critical KV estimation, \sys employs an efficient kernel that fuses the approximation (i.e. MatMul) and selection (i.e. \topk) operations into a single kernel.
The fused kernel updates the \topk selection in-situ without storing the full $QK^T$ tensor with $\mathcal{O}(seq\_len^{2})$ elements, thus reducing memory footprint and computation cost.
Additionally, \sys realizes an efficient sparse attention kernel using query grouping. 
We leverage the observation that queries of adjacent tokens in the latent space often share a significant portion of critical KV pairs.
Their sparse attention computation can then be done together to maximize memory access parallelism and SM utilization.

Third, \sys introduces a sparsity-aware context parallelism (CP) approach. 
Dynamic sparsity breaks CP's basic assumption that each attention head has uniform computation and communication cost due to heterogeneous sparsity across heads. 
\sys adopts a hybrid strategy that performs CP in both head and sequence dimensions: sparse HCP adjusts the head assignment according to per-head sparsity levels to balance the computation load across devices, and sparse SCP further reduces communication overhead by exchanging only the critical KV pairs.
The optimal hybrid configuration is determined by solving an optimization problem that models the effect of both forms of CP.

We build \sys based on PyTorch FSDP~\cite{zhao2023pytorchfsdp} and evaluate it on a testbed with up to 128 H800 GPUs using DiT models ranging from 0.8B to 30B parameters. \sys achieves up to 3.02$\times$ higher training throughput than baseline for input lengths up to 520k, while also reducing end-to-end latency by up to 3.5$\times$. \sys delivers video quality on par with full attention paradigms as also confirmed by human user study.

In summary, our contributions are threefold:

\begin{itemize}[leftmargin=*]
    \item We systematically analyze attention patterns in video DiT training, revealing the existence of critical KV, their unpredictable distribution, heterogeneous sparsity across heads and blocks, and the dynamic evolution of sparsity levels.
    
    \item We propose \sys, a video DiT training framework that leverages dynamic sparsity in attention. \sys integrates adaptive sparse attention computation with specialized kernel and hybrid sparsity-aware context parallelism for optimized performance with multiple devices.
    
    \item We comprehensively evaluate the algorithmic performance and system efficiency of \sys across diverse video generation datasets and video DiT sizes. Results show that \sys maintains quality comparable to full attention while significantly improving throughput and speed.
    
\end{itemize}

\section{Background }
\label{sec:Background}

We start by providing a recap of video DiTs and the general challenges associated with large-scale video DiT training. 

\begin{figure}[t]
  \centering
  \includegraphics[width=0.475\textwidth]{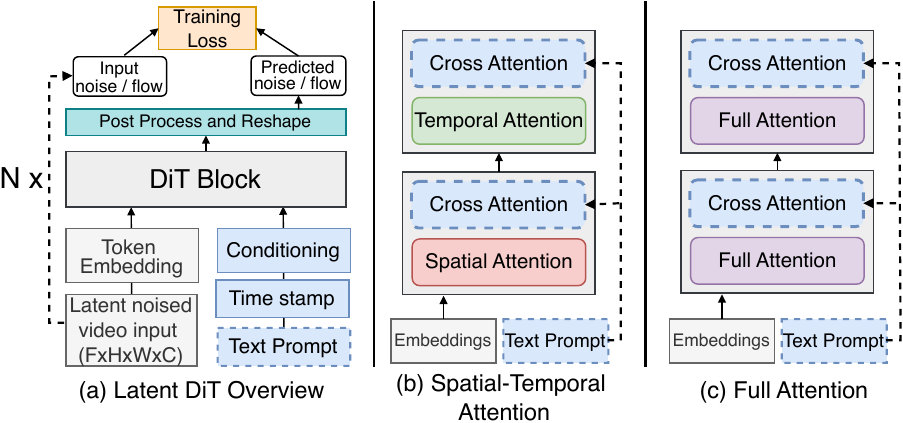}

\caption{Overview of video DiT training. (a) The main input is the video, which is compressed by a VAE (omitted here). The timestamp is used as conditioning, and the text prompt is used as the Key-Value input in the cross-attention module. (b) Interleaved spatial-temporal attention blocks. (c) 3D full attention blocks.}
\label{fig:dit_overview} 
\end{figure}

\subsection{Video DiT} 
\label{sec:video_dit}
Diffusion models have become a leading framework in generative modeling, excelling in image synthesis~\cite{esser2024stablediff3,chen2023pixart} and video generation~\cite{peebles2023scalableDiT,ma2024latte,klingvideo,HunyuanVideo,bao2024vidu,polyak2024movie}. These models corrupt data with noise in a forward process and learn to reconstruct it through a reverse generative process. Among them, DiT has emerged as the de facto backbone backbone~\cite{peebles2023scalableDiT,ma2024latte,OpenSora}.

In a typical video DiT training process, a video clip is first encoded by a VAE~\cite{stabilityai_stable_diffusion,peebles2023scalableDiT,OpenSora}, which compresses and downscales the input into a latent representation. Random noise is added to this latent representation, and the noised latent video-along with any conditioning information (e.g., timestamps or text prompts for text-to-video tasks)-is fed into the DiT model. 
As shown in Figure~\ref{fig:dit_overview}, the DiT model consists of multiple DiT blocks that process video tokens alongside conditional inputs, guiding video generation during training.
At the core of each DiT block are \textit{self-attention} modules, which employ \textit{interleaved spatial and temporal attention} or \textit{full attention} to capture the complex relationships among video tokens across different dimensions. Additionally, \textit{cross attention} aligns the video with text prompts, ensuring consistency between modalities. The model's output is used to compute the loss, following either a denoising diffusion probabilistic model paradigm~\cite{ho2020denoisingddpm} or a flow matching paradigm~\cite{lipman2022flowmatching,ma2024sit}.

\subsection{Various Self-Attention Paradigms} 
\label{sec:self_attention}

Self-attention, or multi-head attention, is widely used in natural language processing and computer vision~\cite{waswani2017attention,dosovitskiy2020imagevit} to capture long-range dependencies. Given an input sequence $H = [h_1, \cdots, h_S]^\top \in \mathbb{R}^{S \times d}$, where $S$ is the sequence length and $d$ the hidden dimension, each attention head maps $H$ to ${Q}$, ${K}$, and ${V}$ using learnable projection matrices $W_Q, W_K, W_V \in \mathbb{R}^{d \times d_k}$. The head output is computed as:
\[
H' = \mathrm{softmax}\Bigl(\frac{\mathbf{Q}\mathbf{K}^\top}{\sqrt{d_k}}\Bigr)\mathbf{V},
\]
where $d_k$ is the dimensionality of each attention head. Ultimately, the outputs from all attention heads are concatenated and combined to produce the final output.

For video DiTs, various attention paradigms have been developed. One approach employs spatial-temporal attention, alternating computations along spatial and temporal dimensions~\cite{ma2024latte}. While computationally efficient, studies have found it inadequate for capturing detailed information, prompting a shift to full attention paradigms~\cite{HunyuanVideo,polyak2024movie,yang2024cogvideox}.

Full attention computes interactions across all tokens in the 3D temporal-spatial space, outperforming spatial-temporal attention. However, it processes a significant number of tokens, often exceeding 100k-even in latent space (e.g., for a latent input of 32$\times$96$\times$96\footnote{Throughout this paper, latent video dimensions (e.g., 16$\times$16$\times$16) are ordered as frames, height, and width unless otherwise specified.}, 300k tokens are required). This makes full attention extremely compute-intensive.

\begin{figure}[t]
  \centering
  \includegraphics[width=0.475\textwidth]{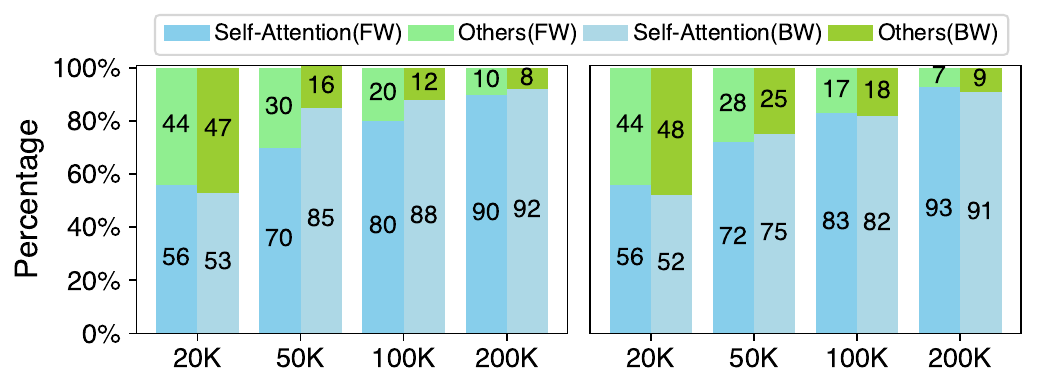} 

  \caption{Time breakdown for self-attention and other operations in different DiTs (left: 1.3B, right: 3B) with various token lengths in forward (FW, left bar) and backward (BW, right bar) computation.} 
  \label{fig:attention_bottleneck} 
\end{figure}

\subsection{Large-Scale Video DiT Training}
\label{sec:dit_training}
The efficiency of large-scale video DiT training is primarily influenced by two key aspects.

\noindent\textbf{Full attention bottleneck.}  In high-resolution, long-video processing, self-attention becomes a major bottleneck due to its $O(n^2)$ complexity with respect to the number of tokens and the non-causal nature of DiT, where the lengths of ${Q}$, ${K}$, and ${V}$ all match the number of video tokens.
In contrast, cross-attention is far more efficient, as only ${Q}$ matches the video token count, while ${K}$ and ${V}$ are constrained to the text prompt length (typically under 120 tokens). As shown in Figure~\ref{fig:attention_bottleneck}, self-attention increasingly dominates computation time as the number of video tokens grows. For example, in 1.3B and 3B models with a sequence length of 200K, self-attention accounts for 92\% and 93\% of forward and backward computation time, respectively.
Thus, developing efficient attention mechanisms is critical to alleviating this bottleneck and enabling scalable video DiT training.

\noindent\textbf{Context parallelism.} Context parallelism (CP) enables long-sequence training by dividing sequences into chunks distributed across multiple GPUs. However, self-attention requires inter-device communication to process the entire sequence, with two common paradigms:

\begin{itemize}[leftmargin=*]
    \item \textbf{Sequence-wise CP.} QKV tensors are split into chunks of shape $[B, H, S/N, D]$, where $B$ is the batch size, $H$ the number of heads, $S$ the sequence length, $N$ the number of GPUs, and $D$ the head dimensionality. Each GPU processes its local query chunk across all heads, gathering KV from other GPUs to compute attention via block-wise operations~\cite{liu2023ring}, often using ring-based communication.

    \item \textbf{Head-wise CP.} GPUs also initially hold QKV chunks partitioned along the sequence dimension for all heads. After All-to-All operations, each GPU gets complete QKV tensors for a subset of heads, reshaped to $[B, H/N, S, D]$. GPUs then compute attention outputs independently for their assigned heads. A final All-to-All re-partitions the outputs to restore the original layout~\cite{jacobs2023deepspeedulysses}. 

\end{itemize}

These paradigms have trade-offs in computation efficiency and communication cost~\cite{gu2024loongtrain,fang2024usp}. Determining an efficient and optimal parallelism strategy is non-trivial, as it depends on the specific input case and hardware configuration.

\section{Empirical Observations on the Sparsity of Attention in Video DiT}
\label{sec:finding}

We conduct an in-depth analysis of the sparsity patterns of attention in video DiT training across various datasets and model sizes, as in~\Cref{sec:eval}. Our findings, demonstrated through a case study, validate the rationale behind our system design. 
The case study focuses on a 2.7B model with an architecture similar to Meta's MovieGen~\cite{polyak2024movie}, representing a moderate scale for video generation~\cite{OpenSora,HunyuanVideo,ma2024latte}. The model is trained on the WebVid-10M dataset~\cite{Bain21webvid_10M} using 8 H100 GPUs with a global batch size of 32 and a latent input size of 16$\times$16$\times$16.

We begin by some definitions.
Given a query vector $q$, let $S_q = \{(k_i, v_i)\}_{i=1}^n$ denote the set of all $n$ key-value (KV) pairs. The attention score function $A(q, k_i) = \text{softmax}(q \cdot k_i)$ computes the scaled dot product between $q$ and each key $k_i$. The set of \textit{critical KV pairs} $I_q \subseteq S_q$ is then defined as the subset of pairs $(k_i, v_i)$ where $A(q, k_i)$ exceeds the $\theta$-percentile threshold of all attention scores.
The threshold $\theta$ can be set by a numerical cutoff or a cumulative sum, such that critical pairs account for 90\% of total attention. In this paper, we use a 90\% cumulative sum threshold as the default, i.e. {\textit{the critical KV pairs are the top ones that together represent 90\% of total attention}}.
The \textit{sparsity} of an attention head is then defined as the average proportion of non-critical KV pairs across all queries: $\mathbb{E}_{q \sim Q}\left[\frac{|S_q \setminus I_q|}{|S_q|}\right]$.

\begin{figure}[t]
    \centering
    \begin{minipage}[t]{0.32\textwidth}
      \vspace{0pt} 
       \centering
   \includegraphics[width=\textwidth]{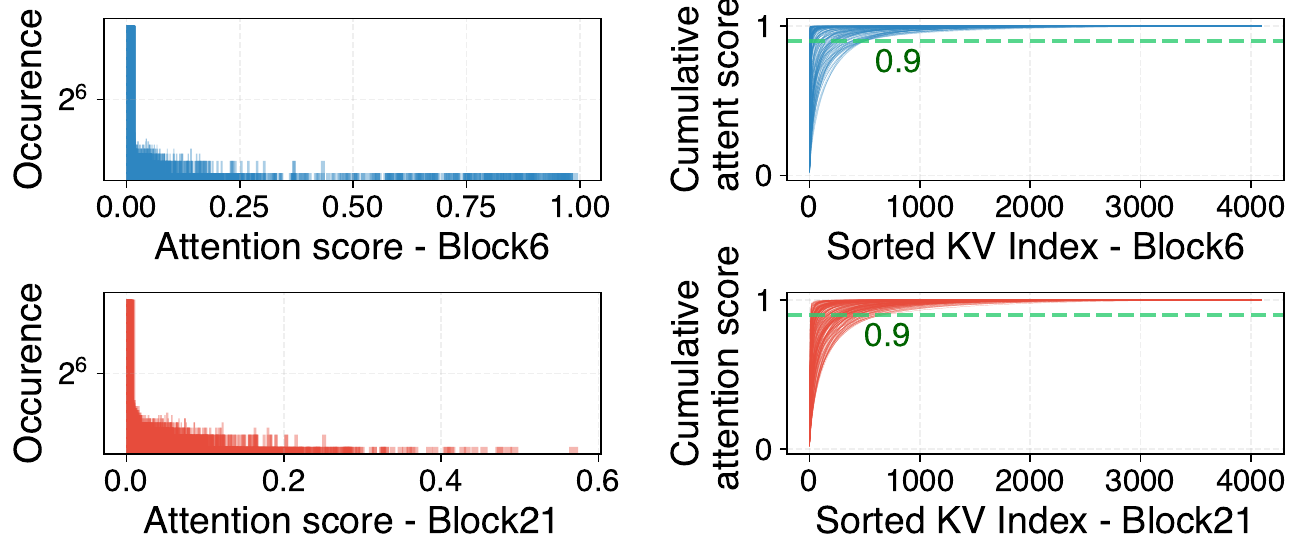}
\caption{Left: The attention score distribution for each query in a histogram. Right: The cumulative distribution function of the sorted attention scores for each query. }
  \label{fig:atten_power_law} 
    \end{minipage}
    \hfill
    \hspace{-0.25in}
    \begin{minipage}[t]{0.15\textwidth}
      \vspace{0pt} 
      \centering
  \includegraphics[width=0.95\textwidth]{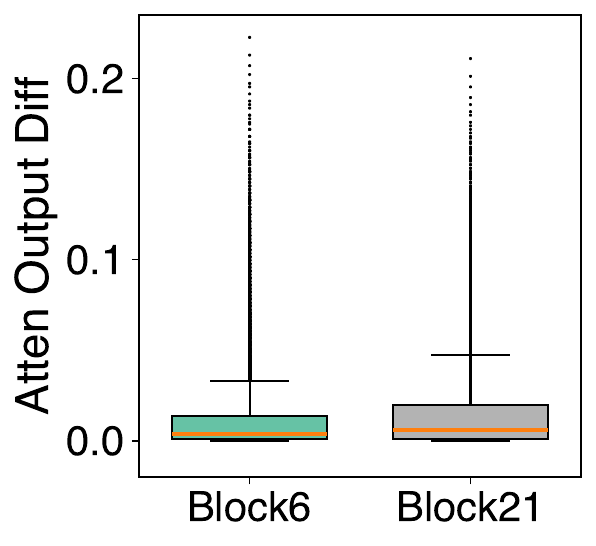}
  \caption{The output difference between attention with full KV and critical KV.}
  \label{fig:output_diff} 
    \end{minipage}
   
\end{figure}

\noindent\textbf{Attention sparsity and power-law distribution.} 
We first investigate the distribution of attention scores for each query to demonstrate attention sparsity.

\hypertarget{obs1}{
\textbf{\textit{Observation 1: Attention scores in sparse blocks follow a power-law distribution.}}}

As shown in Figure~\ref{fig:atten_power_law}, attention scores of queries in two blocks are highly skewed, with most being small (<0.001) and only a few large (>0.1). Notably, the few top keys dominate, with the top 10\% accounting for over 90\% of total scores in 95.2\% of queries (block 6) and 86.8\% (block 21), highlighting the inherent sparsity. Figure~\ref{fig:output_diff} further shows that using only the top 10\% of KV pairs yields minimal output differences.

The power-law distribution of attention scores suggests that a great portion of attention computation may be pruned without significantly impacting performance. By efficiently identifying critical KV pairs, we can potentially reduce computational costs while maintaining model quality.

\begin{figure}[t]
  \centering
  \includegraphics[width=0.475\textwidth]{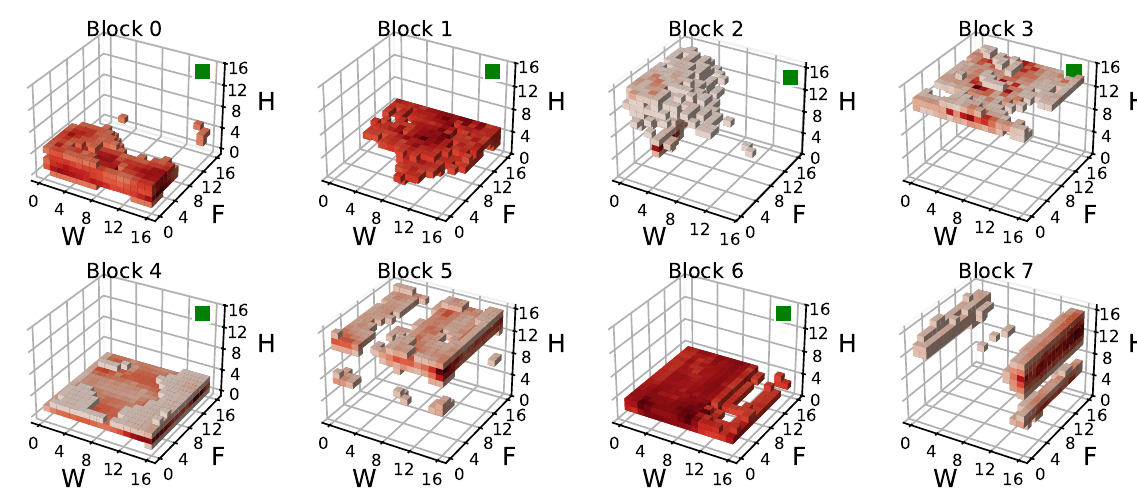}
\caption{Distribution of critical KV positions for a query (green) at position (15, 15, 15) in the 3D latent space. The visualized keys are those that yield attention scores exceeding the 90th percentile when attending to the query.}
    \label{fig:kv_distribution}
\end{figure}

\noindent\textbf{Locality of critical KV.} 
A common and simple method to identify critical KV pairs is to rely on locality: a token's query often attends to nearby tokens' keys or specific global keys, enabling the use of window-based patterns or token sinks as in LLM inference~\cite{xiao2023streamllm,mistral7b,zhang2023h2o}. Given the spatio-temporal dependencies in video DiT, we first investigate whether similar locality patterns exist and can be applied in this context.

\hypertarget{obs2}{
\textbf{\textit{Observation 2: Critical KV pairs do not exhibit locality patterns in video DiT.}}}

In Figure~\ref{fig:kv_distribution}, we visualize the critical KV positions in a 3D space for a query. Contrary to expectations, critical KV pairs do not have a specific pattern, unlike those found in LLM inference~\cite{xiao2023streamllm,mistral7b,zhang2023h2o}. 
More generally across queries, our experiment reveals that only 15.1\% of critical KV pairs are within a 5-token radius, while 48.5\% are more than 10 tokens away. 
This suggests that applying fixed patterns to approximate attention would not fare well in video DiT.

\begin{figure}[t]
  \centering
  \includegraphics[width=0.485\textwidth]{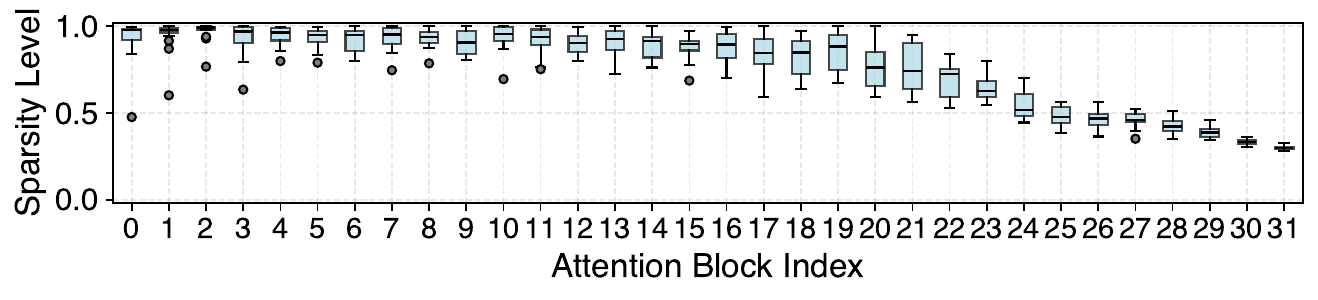} 

\caption{Sparsity of different heads across blocks in iteration 80k. }
  \label{fig:atten_sparsity_dist} 
\end{figure}

\noindent\textbf{Heterogeneity of sparsity.} 
We further establish the spatial heterogeneity of attention sparsity across different attention blocks and heads within a block, which adds to the difficulty of identifying critical KV pairs.

\hypertarget{obs3}{
\textbf{\textit{Observation 3: Sparsity varies significantly across attention blocks and heads within the same block.}}}

\begin{figure}[t]
    \centering
    \begin{minipage}[t]{0.335\textwidth}
       \vspace{0pt} %
       \centering
  \includegraphics[width=\textwidth]{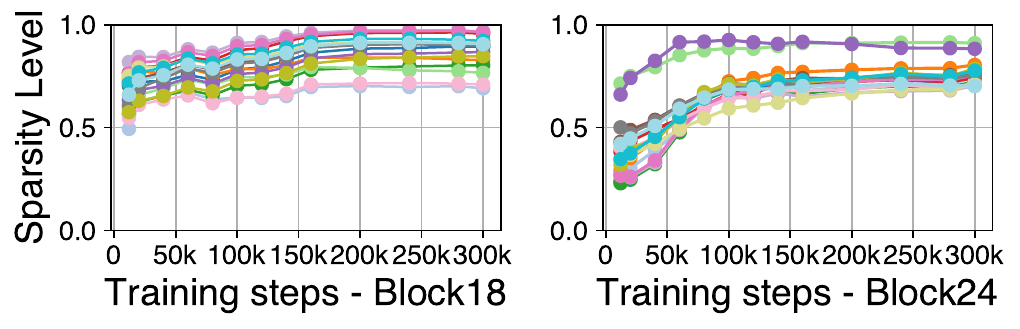}

\caption{The change in sparsity of different attention heads during the training process in transformer blocks. Only two blocks are shown due to space constraints. }
  \label{fig:sparsity_change} 
    \end{minipage}
    \hfill
    \hspace{-0.25in}
    \begin{minipage}[t]{0.135\textwidth}
       \vspace{0pt} %
      \centering
  \includegraphics[width=\textwidth]{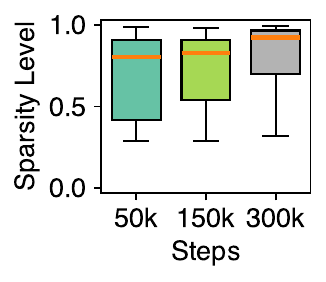} 

  \caption{The sparsity of all blocks across different steps.}
  \label{fig:sparsity_across_blocks} 
    \end{minipage}
   
\end{figure}

Our analysis reveals that sparsity varies not only across blocks but also among heads within the same block. Figure~\ref{fig:atten_sparsity_dist} shows head-level sparsity at iteration 80k, where the first 20 blocks are highly sparse, and latter blocks are less so. The box plot highlights variability within blocks-e.g., in block 2, most heads have 95\% sparsity, with outliers at 90\% and 80\%. Figure~\ref{fig:sparsity_across_blocks} confirms diverse sparsity patterns across blocks at different training steps. Besides, as training converges, \textit{an arch-shaped sparsity pattern} emerges: intermediate blocks become highly sparse, while initial and final blocks go dense.

This disparity aligns with prior studies~\cite{waswani2017attention,jin2024mohattentionmoe,li2019informationattention}, showing that attention heads capture distinct features, with some focusing on local details and others on global context. Uniform sparsity thresholds are suboptimal: low thresholds cause redundant computation, while high thresholds risk losing critical KV pairs for less sparse heads. Adaptive methods are needed to account for each block's and head's unique sparsity, optimizing efficiency and performance.

\noindent\textbf{Time-varying sparsity during training.} 
Continuing from the previous finding, we also observe that attention sparsity varies dynamically in the temporal dimension.

\hypertarget{obs4}{
\textbf{\textit{Observation 4: Sparsity also varies over the course of training before stabilizing.}}}

As illustrated in Figure~\ref{fig:sparsity_change}, sparsity of different heads becomes more pronounced as training progresses. For block 24, the average attention sparsity for each head increases from 0.30 to 0.78. Further, Figure~\ref{fig:sparsity_across_blocks} shows that across all attention blocks, the median sparsity increases from 0.81 at 50k iterations to 0.92 at 300k iterations.
This dynamic evolution of sparsity during training suggests that the strategy for leveraging sparsity should be adjusted dynamically: As the model learns to focus on the most relevant features, the attention mechanism becomes more selective, resulting in increased sparsity~\cite{vig2019analyzingatten}. This highlights the importance of dynamic methods that can capture and exploit the evolving sparsity patterns throughout the training process. 

\begin{figure}[t]
  \centering
  \includegraphics[width=0.425\textwidth]{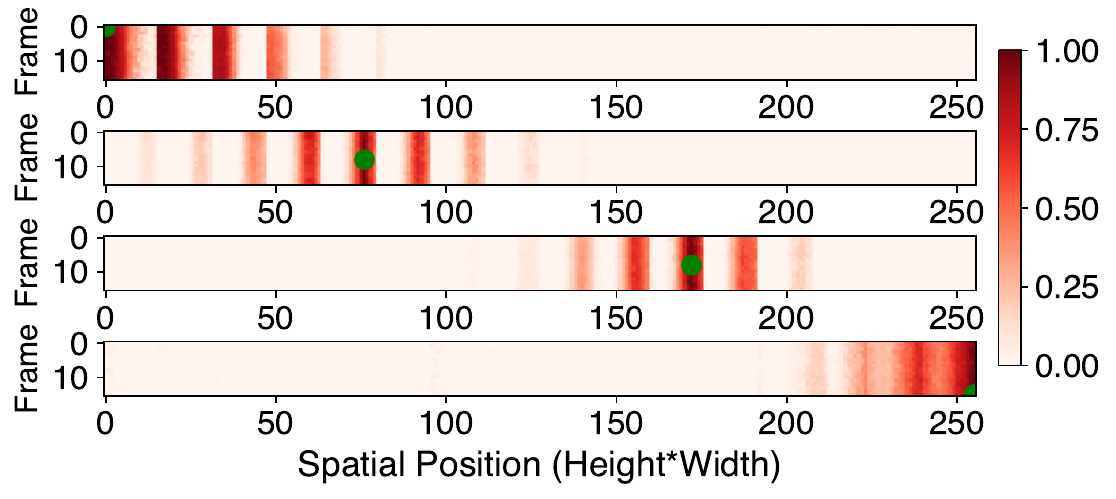}
  \vspace{-0.1in}
    \caption{Critical KV pair overlap ratio between four anchor queries (green, at indices 0, 2124, 2220, 4095) with all queries in an attention block. Each subplot shows a 2D projection of the 3D video with rows representing flattened frames.}
  \label{fig:kv_overlap} 
\end{figure}

\noindent\textbf{Critical KV overlap ratio for adjacent tokens in 3D space.} 
Finally, we demonstrate an interesting phenomenon between queries of adjacent tokens. 

\hypertarget{obs5}{
\textbf{\textit{Observation 5: Adjacent tokens have similar critical KV pairs.}}}

We quantify this phenomenon by measuring the overlap of critical KV indices across queries. Figure~\ref{fig:kv_overlap} is a case study with four anchor queries (in green), where darker shades indicate higher overlap ratios. Significant overlaps in critical KV indices are observed among adjacent tokens. For tokens within a 2$\times$2$\times$2 3D cube (shown as a 2D projection), the four anchor queries and their adjacent queries exhibit over 92.4\% overlap in critical KV indices. On average, this overlap remains consistent at 80.1\% across blocks.
This aligns with the intuition that video pixels represent continuous signals, unlike discrete language signals, making adjacent tokens similar. It suggests neighboring queries tend to attend to similar KV pairs (though these KV pairs may not be close to the query as \hyperlink{obs2}{Obs.2} reveals), offering opportunities to optimize attention computation by leveraging the similarity in sparsity patterns among adjacent tokens.

\section{\sys Overview}
\label{sec:overview}

Our empirical findings in \Cref{sec:finding} uncover inherent sparsity patterns in video DiT attention during training, some of which are different from sparsity in LLM inference for language tasks and have not been previously reported. 
Due to high sparsity, by computing only the most critical KV pairs, the overall computational burden of full attention can be significantly reduced, leading to more efficient training. Moreover, this reduction in KV computations also lowers communication overhead in large-scale settings, where KV data is frequently exchanged for context parallelism. Together, they highlight the potential for substantial end-to-end speedups by exploiting attention sparsity in video DiT training.

Building on this insight, we now discuss the unique challenges posed by these sparsity patterns and introduce the architecture of our solution, \sys.

\subsection{Challenges}
\label{sec:challenges}

Sparsity entails design challenges in three basic aspects of the training system: algorithm, kernel, and parallelism.

\begin{itemize}[leftmargin=*]
\item \textbf{Critical KV identification:} Our first major challenge stems from the dynamic and uncertain distribution of critical KV pairs in video DiT training (\hyperlink{obs2}{Obs.2}). Any fixed sparse attention pattern becomes impractical when the critical KV varies dramatically with changing context and/or training steps. 
While computing the complete attention score matrix (i.e., $\text{softmax}(QK^{T})$) and selecting the \topk entries per query fully capture these changing distributions, it does not save any computation of the attention score.
Moreover, this would disrupt the fused kernels such as FlashAttention~\cite{dao2022flashattention1,dao2023flashattention2} with IO-awareness optimization, limiting the sparsity benefits to only the score-value ($AV$) computation, ultimately degrading overall performance.

\item \textbf{Efficient sparse attention computation:} Even if the attention scores are available for identifying critical KVs, performing \topk selection at scale introduces considerable implementation hurdles. Specifically, the attention score tensor has a shape of $[H, S, S]$, where $H$ is the number of heads and $S$ is the sequence length, which can easily reach up to 100K.
Saving such large tensors alone requires excessive GPU memory.
Furthermore, following top-k selection, each query might access a very sparse set of KV entries in an irregular and scattered pattern. This irregularity can degrade parallel efficiency especially in memory access and result in low use of SM resources, making it difficult to fully reap the gains from sparsity.

\item \textbf{Re-visiting context parallelism:} 
The last challenge emerges when training with long video sequences that must be distributed across devices. 
Existing context parallelism (CP), head-wise and sequence-wise, would fail to work efficiently once sparsity is introduced. 
Head-wise CP can become problematic when certain heads exhibit higher degrees of sparsity than others (\hyperlink{obs3}{Obs.3}), creating load imbalances and stragglers. 
Moreover, standard sequence-wise CP often transfers all KV pairs among devices without considering which ones are critical~\cite{liu2023ring}, unnecessarily increasing communication overhead. As a result, we need to overhaul the parallelization strategy to account for the idiosyncrasies due to sparsity.

\end{itemize}

\begin{figure}[t]
    \centering
    \includegraphics[width=0.405\textwidth]{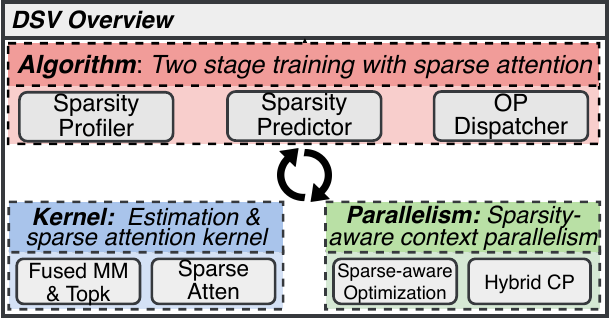} 
  \caption{\sys overview.}
    \label{fig:sys_overview} 
\end{figure}

\subsection{\sys Architecture and Workflow}
\sys addresses the three challenges with three key designs as summarized in Figure~\ref{fig:sys_overview}.

\begin{itemize}[leftmargin=*]
  \item {\bf \em Algorithm: Two-stage training with low-rank $QK$-estimation.} At the core of \sys lies an efficient low-rank estimation of query and key to approximate the attention score. Two low-rank matrices are trained for each attention module to approximate $QK^T$ (separate from FlashAttention). These matrices enable identification of critical KV pairs without fully computing attention scores or modifying FlashAttention kernels.
  DiT training then becomes a two-stage process: in the first stage these low-rank matrices, called \textit{sparsity predictors} hereafter, are continuously learned until their approximation is accurate enough, and all attention is computed in full. 
  Then we enter the second stage: in each step, when the sparsity level exceeds a threshold, we perform sparse attention over the critical KV pairs found by the low-rank estimations.

  \item {\bf \em Kernel: Critical KV estimation and sparse attention.}  To overcome the second challenge and maximize the benefits of sparse training, \sys introduces optimized kernels tailored for critical KV estimation and sparse attention. Specifically, a fused kernel calculates the approximate attention scores using learned low-rank matrices and performs top-$k$ selection at the desired sparsity level in a single pass to minimize intermediate memory consumption.
  Additionally, \sys adopts a sparse attention kernel with query grouping to enhance memory access and computation parallelism, as queries that are close in position often share similar critical KVs as noted in \hyperlink{obs5}{Obs.5}.
  
  \item {\bf \em Parallelism: Sparsity-aware context parallelism.}  
  Lastly, \sys leverages sparsity-aware context parallelism (CP) strategies to dynamically adapt to varying sparsity levels across different attention blocks and heads. It introduces new optimizations for sparse settings, including head-wise workload reallocation for standard head-wise CP and selective KV gathering for seq-wise CP. Further, \sys determines the optimal hybrid parallelism configuration for each attention block based on its head-wise sparsity patterns. By jointly optimizing computational load balancing (mitigating inefficiencies from sparsity-induced workload imbalances) and communication overhead across paradigms, it ensures optimal performance when parallelizing large video inputs.
\end{itemize}

\section{Two-Stage Training with Low-Rank Based Sparsity Prediction}
\label{sec:algorithm}

\begin{figure}[t]
  \centering
  \includegraphics[width=0.475\textwidth]{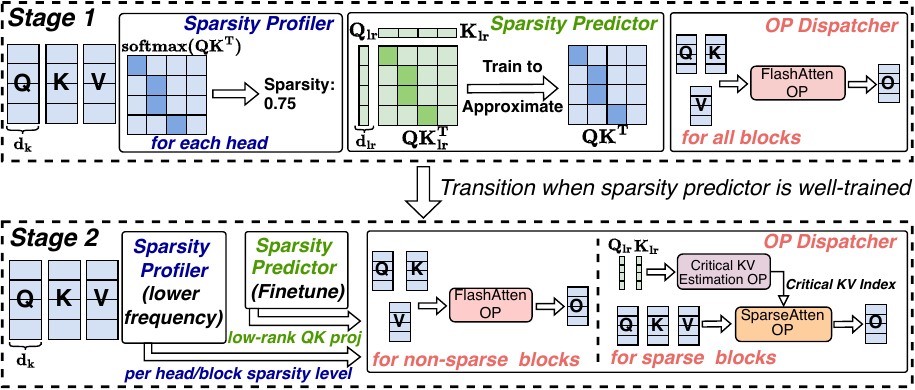} 
  \caption{Two-stage DiT training paradigm.} 
  \label{fig:algorithm} 
\end{figure}

We now present \sys's training algorithm design as depicted in Figure~\ref{fig:algorithm}.

\subsection{Sparsity Profiling}
\label{sec:profiling}
We first introduce \sys's profiler as the basic building block for both low-rank based sparsity prediction and two-stage training.
The profiler periodically measures the sparsity level of each attention head as defined in~\Cref{sec:finding}, by separately computing the softmax of $QK^T$ without disrupting the normal full attention computation using FlashAttention. 
To reduce the overhead, random sampling (e.g., by a factor of 16) on query $Q$ is used when examining each head's attention scores. 
At iteration $i$, we update the exponential moving average of sparsity level of the current block $S^{i}$ with the latest sample $P^{i}$: $
S^{i} = \alpha  P^{i} + (1 - \alpha) S^{i-1} $,
where $\alpha$ is the smoothing factor that smooths out fluctuations due to sampling noise.
The profiler maintains sparsity level for each attention head and block, which are used to train the low-rank query and key matrices and to determine the use of sparse computation.

\subsection{Low-Rank based Sparsity Prediction}
\label{sec:lowrank}
For each attention block, we introduce two low-rank trainable matrices, $W_Q^{\text{lr}}$ and $W_K^{\text{lr}}$, which project the input to $Q_{\text{lr}}$ and $K_{\text{lr}}$ with an inner dimension $d_{\text{lr}}, d_{\text{lr}} \ll d_k$ (default: 16) where $d_k$ is the inner dimension of original query and key, $Q$ and $K$. 
We train $W_Q^{\text{lr}}$ and $W_K^{\text{lr}}$ at \textit{each} iteration of DiT training such that $Q_{\text{lr}}K_{\text{lr}}^T$ can \textit{continuously} approximate $QK^T$ from the profiler. 
Note this predictor training process is independent from the primary computation graph. 
It ensure accurate approximation of attention patterns with minimal parameters and overhead (<10M for a 3B model).

We train the sparsity predictor for each block using the following loss function: $0.95\cdot \text{CosLoss}(Q_\text{lr}K_{\text{lr}}^{T}, QK^{T}) + 0.05\cdot \text{NormLoss}(Q_\text{lr}K_{\text{lr}}^{T}, QK^{T})$.
It encourages the low-rank projections to preserve the relative magnitudes of $QK^T$ elements while maintaining consistency with the original $QK^T$. During the forward pass, the sparsity predictor computes the approximation loss and immediately updates its parameters, reducing memory consumption by avoiding storage of intermediate $QK^T$ results. Random sampling on query $Q$ is also used to further rein in the overhead.

\subsection{Two-Stage DiT Training}
\label{sec:two_stage}

DiT training in \sys proceeds in two distinct stages to ensure effective exploitation of attention sparsity while maintaining model performance.

\noindent\textbf{Stage 1: Full Training.}
In the initial stage, the DiT model undergoes full training without sparse computations. The primary goal is to train the sparsity predictors for each attention block as said before. 
This stage concludes once the average approximation loss across all blocks falls below a defined threshold (default: 0.01). Typically, this condition is met within 5K iterations, signaling that the sparsity predictors are adequately trained and ready for use.

\noindent\textbf{Stage 2: Sparse Training.}
In Stage 2, sparse attention is utilized whenever possible to fully accelerate DiT training. 
The key here is to determine whether to activate sparse computations, which is precisely what \sys's OP Dispatcher is responsible for. 
The signal it uses is obviously the sparsity level: the profiler now runs with a lower frequency than that in Stage 1 due to the stabilized sparsity patterns to monitor each block's sparsity level.
Then whether or not sparse attention is beneficial given a certain sparsity level hings on two factors, the computation speedup and the memory overhead.
Computation speedup includes the gain from sparse attention minus the additional overhead from critical KV estimation using the low-rank sparsity predictor.
Memory overhead also arises from the need of storing the critical KV indices.
To efficiently navigate this tradeoff, we perform offline experiments to measure at various sparsity levels the corresponding speedups and memory requirements (using our re-designed kernels in \Cref{sec:kernels}) for varying video token lengths.
Then OP Dispatcher enables sparse attention for a block when (1) the current memory utilization is sufficient for the corresponding memory overhead, and (2) the current sparsity level exceeds the threshold established offline for this setting.
This block's low-rank sparsity predictor is used to locate the critical KV pairs, and sparse attention is performed using our optimized kernels.

\begin{figure}[t]
  \centering
  \includegraphics[width=0.45\textwidth]{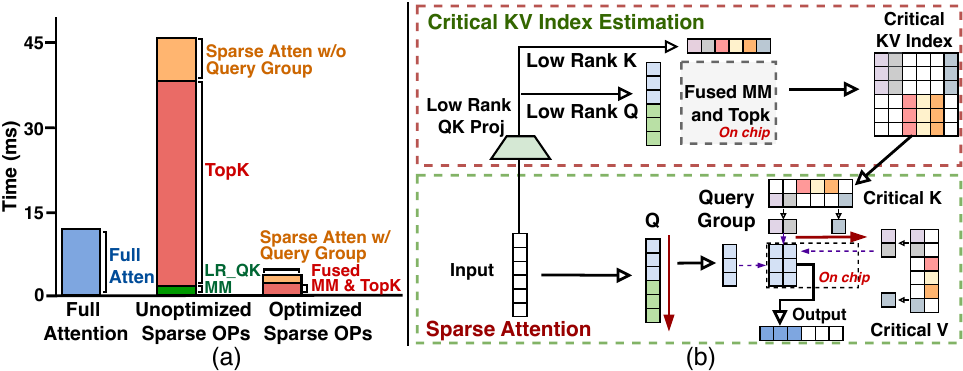} 
    \caption{(a) Full vs. sparse attention operations (head size: 1, token length: 51200, sparsity: 90\%, on H100). (b) Kernel overview in \sys.}
  \label{fig:atten_kernel_overview} 
\end{figure}

\section{Efficient Kernels}
\label{sec:kernels}
This section tackles the challenges of sparse attention implementation, including: (1) critical KV estimation via top-K selection from $Q_{\text{lr}}K_{\text{lr}}^T$, and (2) hardware-aware sparse attention kernels with selected KV pairs. 

\subsection{Critical KV Estimation with Kernel Fusion}
\label{sec:estimation_kernel}

\noindent\textbf{Memory constraints.} 
The critical KV estimation involves low-rank matrix multiplication $Q_{\text{lr}}K_{\text{lr}}^T$ for approximating $QK^T$, followed by a \topk selection operation. 
They present two major bottlenecks: (1) Storing the entire $Q_{\text{lr}}K_{\text{lr}}^T$ demands an excessive memory footprint.
When \(H=16\) attention heads and \(S=100k\) video tokens are used, the resulting \([H, S, S]\) tensor requires $\sim$320GB memory in BF16.
(2) \topk is memory-bound, saturating GPU memory bandwidth for large inputs and significantly reducing overall throughput.
In Figure~\ref{fig:atten_kernel_overview}, \topk takes longer than full attention, consuming up to 80\% of the total naive sparse attention time. 

\noindent\textbf{Kernel fusion.} 
To reduce memory footprint and data movement overhead, we fuse \topk directly into the low-rank MatMul. Instead of materializing the large result tensor, partial MatMul results are immediately used for incremental \topk updates which is all we need here after all.
A custom GPU kernel interleaves these steps, keeping only the \topk entries per query in registers. This reduces the space complexity from \(\mathcal{O}(S^2)\) to \(\mathcal{O}(S K)\) and minimizes data movement between HBM and on-chip registers.

\subsection{Sparse Attention with Query Grouping}
\label{sec:sparse_kernel}
Implementing sparse attention naively where each query relies on distinct critical KV pairs yields limited speedups (about 1.4$\times$ in Figure~\ref{fig:atten_kernel_overview}), primarily due to uncoalesced memory access patterns and diminished tensor core utilization. Since each query can fetch disjoint KV entries, data reuse decreases dramatically, hindering parallel efficiency.
As noted in \hyperlink{obs5}{Obs.5}, adjacent queries in the 3D spatial-temporal domain often share a large portion of critical KV pairs. 
This allows us to cluster nearby queries into groups based on their proximity in 3D cubes (e.g., 2$\times$2$\times$2 voxels), and share the critical KV indices within a group. 
We employ an adaptive offline profiling mechanism to accommodate variations in image resolution and frame rates across datasets and balance hardware efficiency. This mechanism determines the largest feasible group size that satisfies the required overlap ratio (e.g., 80\%) of critical KV for queries within a group. The central query in the group is used exclusively as the proxy to identify the shared critical KV pairs (reduce the estimation problem size as well). Then, these queries are processed together for their attention to improve memory access and computation efficiency.

\section{Hybrid Sparsity-Aware Context Parallelism}
\label{sec:parallelism}
Sparsity in attention adds a new dimension to context parallelism (CP). In this section, 
{we first analyze the performance of head-wise and sequence-wise CP in sparse settings with the best tuning. Building upon this, we propose a hybrid sparse CP strategy that computes the best configuration by jointly optimizing both forms of CP for each attention block.}

\subsection{Modeling Context Parallelism with Sparsity}
\label{subsec:parallel_modeling}
We consider $N$ GPUs within a CP group, each initially processing a partial sequence and storing the corresponding $Q$, $K$, and $V$ chunks with shape [$H$, $S/N$, $D$], where $H$ is the number of heads, $S$ sequence length, and $D$ head dimension.

\subsubsection{\bf \em Head-wise CP with Sparsity}

\begin{figure}[t]
  \centering
  \includegraphics[width=0.475\textwidth]{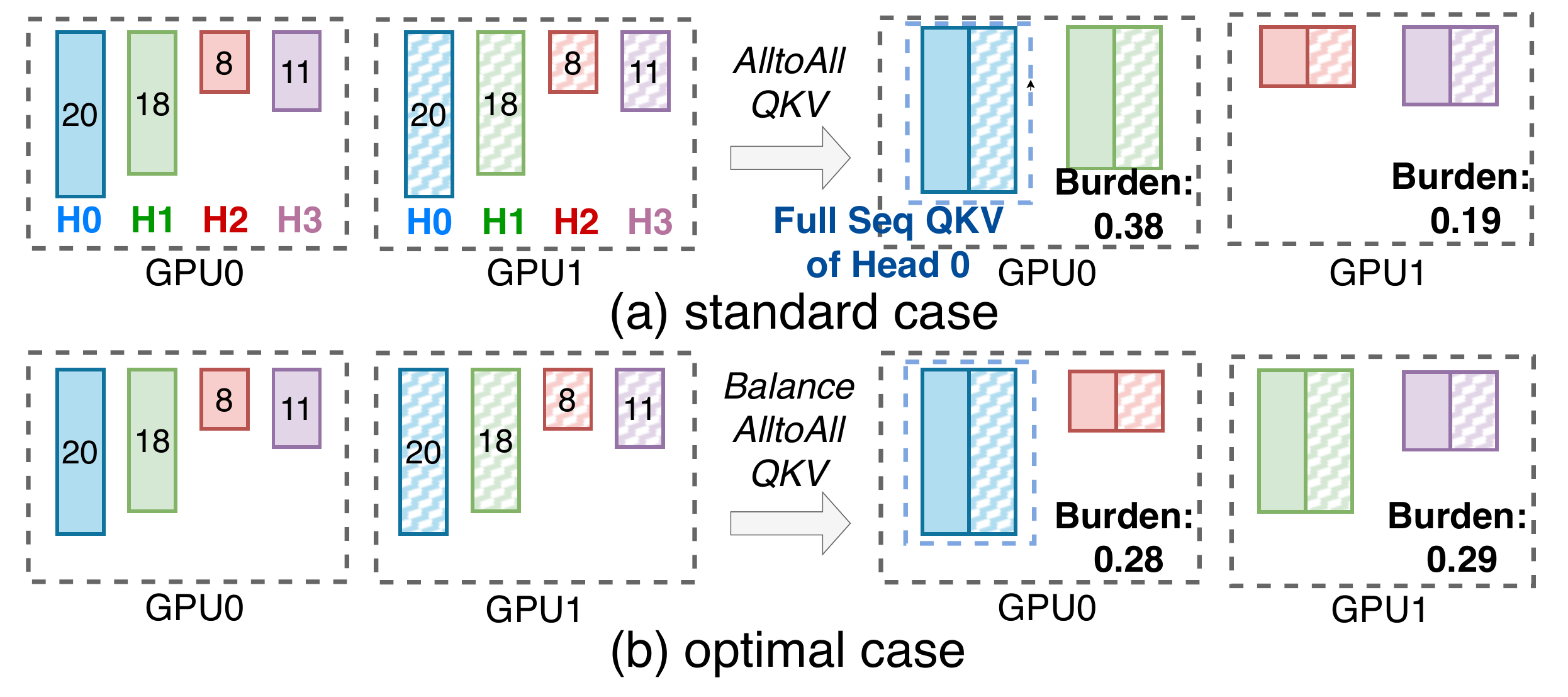} 
  \caption{An example of load imbalance in HCP with four attention heads (different colors). A sequence is split into two chunks (different shades) each held by one GPU. A chunk's length on a head represents its relative computational burden, i.e. 1 minus the sparsity level. A GPU's total burden is the sum of that across all heads it hosts.}
  \label{fig:hcp} 
\end{figure}

HCP redistributes the attention workload across the head dimension using all-to-all operations. Each GPU transitions from processing a sequence partition across all heads to handling a few heads for the entire sequence. However, the varying sparsity levels across heads introduce additional complexity and inefficiency.

\noindent\textbf{Computation load imbalance.} The head-wise split for attention computation, combined with sparsity heterogeneity leads to uneven computational burdens across GPUs as shown in Figure~\ref{fig:hcp}. Here, attention heads are distributed uniformly across GPUs, resulting in a significant load imbalance: GPU0 becomes a straggler due to low sparsity levels of its heads. Swapping the head-GPU assignment reduces the end-to-end time by 35.7\% with more balanced workloads. This highlights the need for a {\textit{balanced head-wise reallocation}} to account for sparsity heterogeneity. 
The best head allocation scheme within a HCP group can be found by a small combinatorial optimization problem that minimizes the maximum computational burden per GPU. We refer to this minimal per-GPU computation burden as $comp^{hcp}$.

\noindent\textbf{Communication cost.} 
We turn to analyze the total communication volume of each GPU in a sparse HCP group. 
Let $H_i^{r}$ represent the final number of heads allocated to GPU ${i}$. The training process involves four uneven all-to-all operations ($QKV$ and output) for each GPU, with total cost amounting to:
$comm_i^{hcp}=3SD\max(H_i^{r}(N-1)/N , (H-H_i^{r})/N ) + SD\max((H-H_i^{r})/N, H_i^{r}(N-1)/N )$.

\noindent\textbf{Memory cost.} 
Similar to communication, the memory consumption for $QKV$ and output on each GPU of a HCP group could slightly differ due to the potential uneven head allocation, and is given by: {$mem_{i}^{hcp}=4 S D H_i^{r}$}.

\subsubsection{\bf \em Sequence-wise CP with Sparsity} 
SCP does not reallocate the computation burden. Instead, it computes the attention output for the local $Q$ with both local $KV$ and remote $KV$ gathered from other GPUs within the same group.

\begin{figure}[t]
  \centering
  \includegraphics[width=0.45\textwidth]{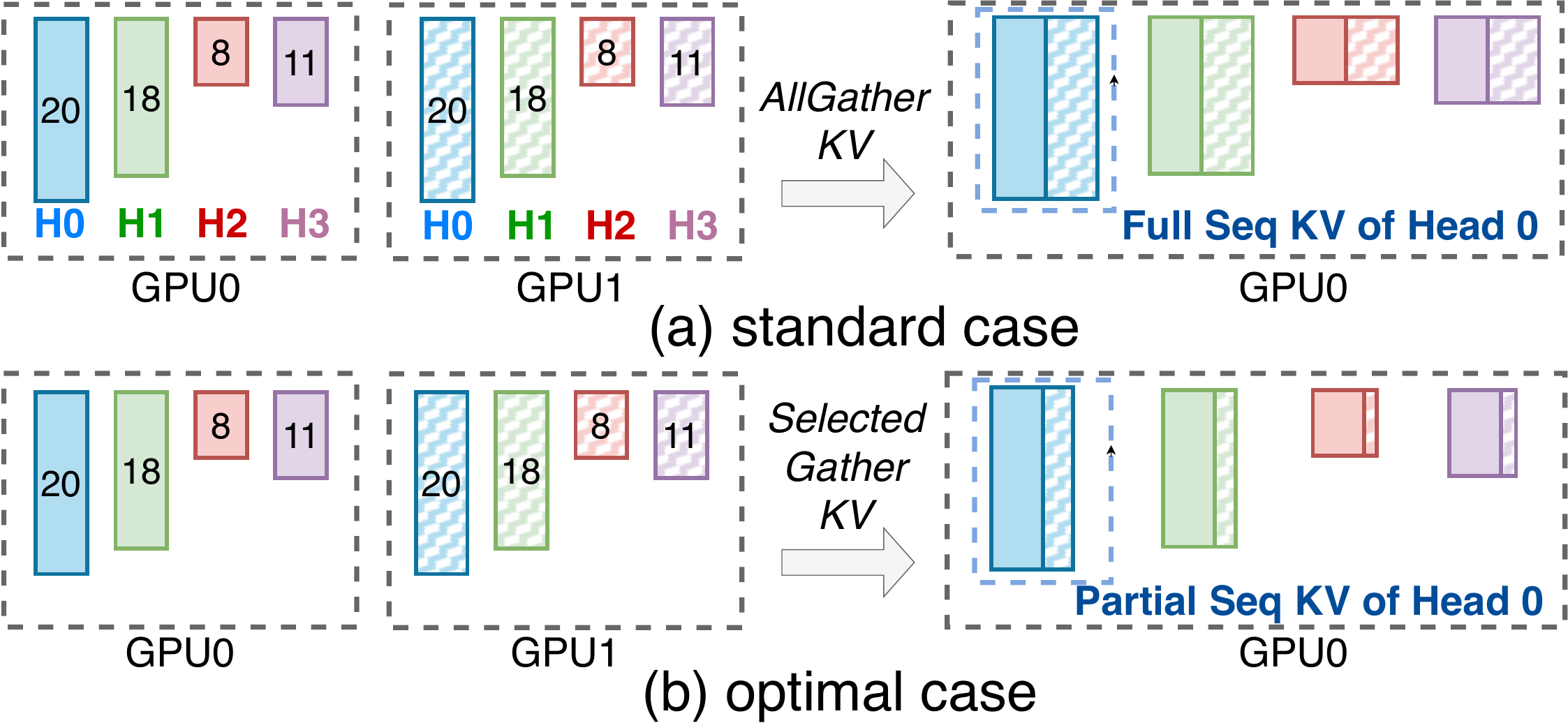} 
    \caption{An example shows redundant communication of SCP in the same setting as in Figure~\ref{fig:hcp}. The width of each chunk represents the held KV size for that chunk. For simplicity, only GPU 0 is shown here after the KV gathering; GPU 1 should have a similar situation.}
  \label{fig:scp} 
\end{figure}

\noindent\textbf{Communication cost.} 
In standard SCP a GPU gathers all remote KV tensors, but this is unnecessary in the sparse setting as shown in Figure~\ref{fig:scp}(a). Only the critical KV tensors are needed now, which leads to \textit{selective KV gathering} as in Figure~\ref{fig:scp}(b). 
Consequently, the actual communication volume is:
{$comm_i^{scp}=2HDS/N\cdot \max(\sum_{j\in N,j\neq i}\alpha_{i}^{j}, \sum_{j\in N,j\neq i}\alpha_{j}^{i})$}, 
where $\alpha_{i}^{j}$ denotes the fraction of KV tensors on GPU $j$ that is needed by GPU $i$.

\noindent\textbf{Memory cost.} 
Similarly, the memory consumption for additional non-local critical KV varies with $\alpha_{i}^{j}$ as well. It can be expressed as:
{ $mem_i^{scp}=2HD\sum_{j\in N,j\neq i}(\alpha_{i}^{j}S/N).$}

\subsection{Hybrid Sparse Context Parallelism } 

The analysis of optimized HCP and SCP in sparse settings highlights their respective strengths and limitations. 
HCP has a big impact on each GPU's computation load, while SCP does not. However, the communication overhead for both varies depending on the sparsity pattern.

Naturally, HCP and SCP can be combined with HCP re-balancing the workloads across heads and SCP further splitting them across sequences for certain heads. For example, in an 8-GPU setup {with a HCP degree of 4,} ranks 0-3 and 4-7 form HCP groups where head-wise reallocation occurs. Each GPU pair across HCP groups (e.g., rank 0 and rank 4) then forms four SCP groups (SCP degree 2), with ranks in each SCP group handling different sequence chunks for the same heads. 
By flexibly partitioning workloads across both dimensions, the hybrid approach (see Algorithm~\ref{algo::cp} in Appendix~\ref{appendix:algo}) can potentially achieve an optimal balance of computation and communication under certain sparsity pattern.

\noindent\textbf{Optimal CP configuration.} 
To find the best configuration for hybrid CP, we formulate an optimization problem that arrives at the optimal HCP and SCP degrees $g^{h},g^{s}$ for each attention block:
\begin{align}
 \min\ & \mathop{\max}\limits_{i}\left({T_{i}^{comm}(g^{h},g^{s})}+{T_{i}^{comp}(g^{h},g^{s})}\right) \nonumber\\
\text{s.t.}\ \ & T_{i}^{comp}(g^{h},g^{s})=f_i(comp^{hcp}(g^{h})), \label{cons:comp} \\
& T_{i}^{comm}(g^{h},g^{s})=h_i(comm_i^{hcp}(g^{h})+comm_i^{scp}(g^{s})), \label{cons:comm} \\
& mem_i\left(g^{h},g^{s}\right) = mem_i^{hcp}(g^h)+mem_i^{scp}(g^s) \leq M, \label{cons:mem} \\
& g^{h} \cdot g^{s} = N,\;  g^{s}\ge 1, \; 1 \leq g^{h} \leq H.\label{cons:group} 
\end{align}
The objective is to minimize the maximum execution time, including both communication and computation on any GPU given the CP configuration $g^{h},g^{s}$. 
The computation time depends on the minimum per-GPU computation burden $comp^{hcp}$ found for $g^{h}$ in cons.~\eqref{cons:comp}, while the communication time is directly determined by the communication volume in cons.~\eqref{cons:comm} (and intra- and inter-node bandwidths which are known).
Cons.~\eqref{cons:mem} ensures that memory usage remains within acceptable limits, while \eqref{cons:group} imposes restrictions on group sizes: the HCP group size is bounded by the model's head count, and the product of the HCP and SCP group sizes must equal the total number of GPUs. 
Given the limited search space and the stable sparsity patterns over training intervals, the optimization can be periodically solved with small overhead.
For deployment, we dynamically determine whether to prioritize HCP-first or SCP-first intra-node placement based on their estimated communication volume under a specific configuration and sparsity pattern.

\section{Implementation}

We implement \sys on PyTorch's FSDP framework, extending it for tensor and context parallelism. The sparse attention kernel is built with Triton~\cite{openai_triton}, and critical KV estimation in CUDA. \sys provides a non-invasive module that integrates seamlessly with any training framework and DiT model. Its low-rank sparsity prediction is decoupled from the main model training, allowing replication across devices and manual control of computation and gradient synchronization.
Other optimizations such as CPU offloading for large KV indices are in Appendix~\ref{appendix:imple}.

\section{Evaluation}\label{sec:eval}

\noindent\textbf{Setup.} We evaluate on servers with 8 NVIDIA H800 GPUs connected via NVLinks and 200 Gbps RoCE for cross-node communication. 
Training is done in BF16 except for gradient reduction and optimizer updates which use FP32.

\noindent\textbf{Datasets.} We adopt four datasets: the widely used UCF-101~\cite{soomro2012ucf101} and Webvid-10M~\cite{Bain21webvid_10M} by prior work, and the more recent VideoGen~\cite{tan2024vidgen} and OpenVid~\cite{nan2024openvid} that feature high-definition videos with detailed text descriptions.

\noindent\textbf{Models.} The model configurations are listed in Table~\ref{table:model_config}. Their architectures, based on DiT blocks with 3D self-attention and cross-attention, are similar to Meta MovieGen~\cite{polyak2024movie} and other SOTA models~\cite{klingvideo,HunyuanVideo}. For video compression, we use Stability-AI's VAE~\cite{stabilityai_stable_diffusion} to reduce spatial resolution by 8$\times$8. The CLIP text encoder~\cite{clip_textembedder} is used for UCF-101 and WebVid-10M, while the T5-xxl text encoder~\cite{T5encoder} is applied to VideoGen and OpenVid. Models are trained with the Adam optimizer (learning rate: 1e-4) and gradient checkpointing~\cite{chen2016gradcheckpoint} is enabled. Flow matching~\cite{lipman2022flowmatching}, a widely adopted generative modeling approach, is used for its advanced performance~\cite{OpenSora,HunyuanVideo,polyak2024movie}.

\noindent\textbf{Baselines.} We compare \sys with the following baselines: %

\begin{itemize}[leftmargin=*]
    \item \textbf{Vanilla full attention (FA).} The 3D full attention computation based on FlashAttention-2~\cite{dao2023flashattention2,tritonflashattention} is the de facto implementation choice today.

    \item \textbf{Window-based attention (WA).} A common way to use sparse attention in video models~\cite{hassani2023neighborhood,zhang2025fastvideotileatten} where a query attends to KV pairs in a 3D window centered on itself. We explored WA-Medium ({WA-M}) with each dimension at 1/3 of the original size, and WA-Large ({WA-L}) at 2/3 instead.
\end{itemize}

For context parallelism, we prioritize head-wise CP for baselines due to its superior performance in our testbed. If the CP degree exceeds the number of head, we adopt intra-node head-wise CP with inter-node sequence-wise CP.

\begin{table}[t]
\resizebox{\columnwidth}{!}{%
\begin{tabular}{cccccc}
\hline
Model & \#Layer & \#Head & Head size & Activation & Norm Type \\ \hline
0.8B & 28 & 12 & 96 & GeGLU & AdaLayerNormSingle \\
2.7B & 32 & 16 & 128 & GeGLU & AdaLayerNormSingle \\
30B & 42 & 24 & 256 & GeLU-approximate & AdaLayerNormSingle \\ \hline
\end{tabular}%
}
\caption{Model configurations in our evaluation.}
\label{table:model_config}
\vspace{-4mm}
\end{table}

\begin{figure*}[t]
  \centering
  \includegraphics[width=0.98\textwidth]{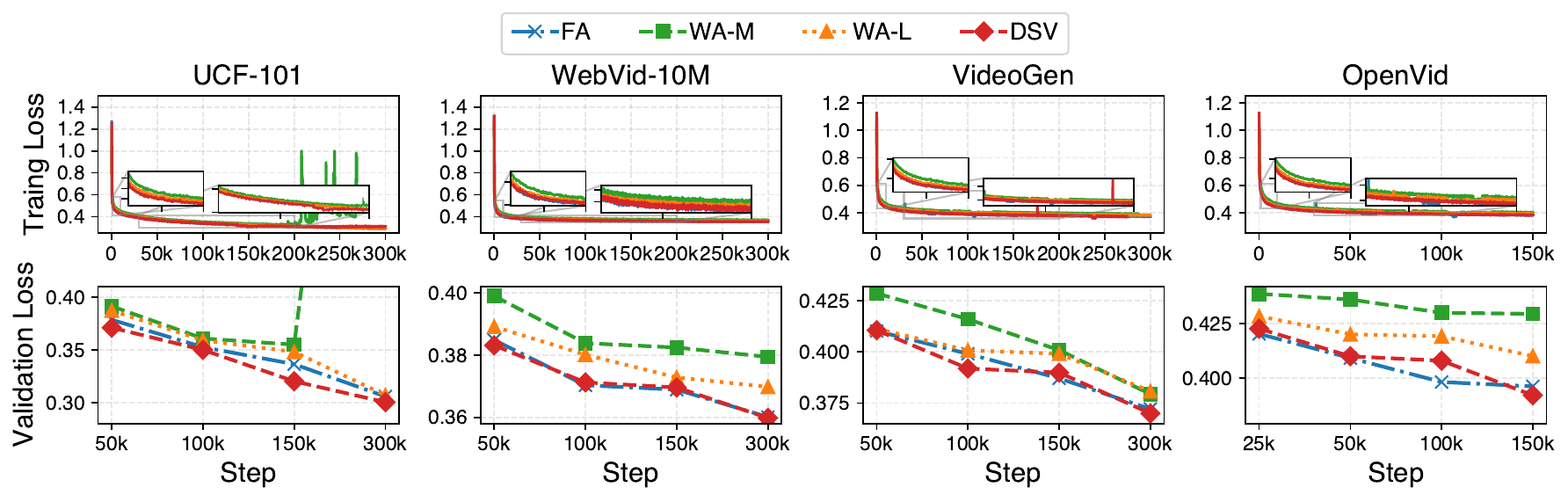} 
\caption{Comparison of training loss and validation loss throughout the training process for different methods across four datasets.}
  \label{fig:losses} 
\end{figure*}

\subsection{Overall Model Quality}
\label{sec:quality}
We compare \sys's training convergence and model quality with baselines across various configurations. 
A small 0.8B model is trained on UCF-101 and WebVid-10M with a latent input size of 16$\times$16$\times$16 (4k tokens), while bigger 2.7B models are trained on VideoGen and OpenVid with latent input of 32$\times$32$\times$32 (32k tokens) and 40$\times$56$\times$56 (125k tokens), respectively. 

\noindent\textbf{Training and validation loss.} Figure~\ref{fig:losses} compares the training and validation loss trends across different attention mechanisms. 
These losses reflect model improvement in the flow matching paradigm~\cite{polyak2024movie}. 
\sys exhibits faster convergence and achieves lower final loss values compared to WA-M and WA-L, and is comparable to FA across all datasets and models. Notably, WA-M fails to converge {on UCF-101}. Moreover, for validation loss, \sys also closely matches FA while outperforming both WA methods.

\noindent\textbf{Video quality.} Table~\ref{table:quality_comp} presents the video generation quality of \sys compared to other baselines across various datasets and evaluation metrics. 
Consistent with prior work, we use FVD~\cite{unterthiner2018fvd} to quantify differences between generated and original videos, supplemented by VBench~\cite{huang2024vbench} which measures quality and semantic coherence through model-based methods.

In terms of FVD, \sys consistently achieves the lowest values or performs comparably to FA, indicating its ability to generate videos that closely resemble the ground truth (FA). In contrast, WA-M consistently exhibits poor performance. Similarly, for semantic and quality scores in VBench, \sys continues to demonstrate superior performance.

\begin{table}[t]
\centering
\begin{minipage}[t]{0.26\textwidth} %
\vspace{0pt} %
\resizebox{\columnwidth}{!}{%
\huge
\begin{tabular}{lcccc}
\hline
 &  & \multicolumn{3}{c}{\textbf{Quality}} \\ \cline{3-5} 
\textbf{Dataset} & \textbf{Method} & \textit{FVD $\downarrow$} & \textit{Quality $\uparrow$} & \textit{Semantic $\uparrow$} \\ \hline
\multirow{4}{*}{UCF-101} & WA-M & 700.98 & 69.18\% & 55.03\% \\
 & WA-L & 520.92 & 70.70\% & 55.84\% \\
 & FA & 440.32 & 71.00\% & 56.23\% \\ \cline{2-5} 
 & \sys & 438.02 & 71.08\% & 56.60\% \\ \hline
\multirow{4}{*}{WebVid} & WA-M & 580.12 & 71.10\% & 40.10\% \\
 & WA-L & 530.38 & 73.23\% & 40.56\% \\
 & FA & 409.24 & 73.79\% & 42.56\% \\ \cline{2-5} 
 & \sys & 414.56 & 73.66\% & 42.88\% \\ \hline
\multirow{4}{*}{VideoGen} & WA-M & 1395.23 & 77.33\% & 53.08\% \\
 & WA-L & 1100.72 & 78.82\% & 53.32\% \\
 & FA & 908.91 & 79.39\% & 55.34\% \\ \cline{2-5} 
 & \sys & 834.32 & 79.14\% & 54.99\% \\ \hline
\multirow{4}{*}{OpenVid} & WA-M & 884.15 & 78.78\% & 55.98\% \\
 & WA-L & 826.43 & 79.47\% & 56.51\% \\
 & FA & 838.52 & 79.54\% & 56.07\% \\ \cline{2-5} 
 & \sys & 782.22 & 79.63\% & 56.36\% \\ \hline
\end{tabular}%
}
\caption{Comparison of model quality metrics on four datasets.} 
\label{table:quality_comp}
\end{minipage}%
\hfill %
\begin{minipage}[t]{0.2\textwidth} %
\vspace{0pt} %
\resizebox{\columnwidth}{!}{%
\begin{tabular}{lcccc}
\hline
      & FA   & WA-M & WA-L & \sys  \\ \hline
\textbf{Score}$\uparrow$ & 4.25 & 2.89 & 3.81 & 4.57 \\ \hline
\end{tabular}%
}
\caption{Normalized user rating scores (1-5).}
\label{table:user_study}

\resizebox{\columnwidth}{!}{%
\huge
\begin{tabular}{cccccc}
\hline
\textbf{}                   & \textbf{}             & \textbf{} & \multicolumn{3}{c}{\textbf{Method}} \\ \cline{4-6} 
\textbf{Dataset} & \textbf{Model} & \textbf{Length} & \multicolumn{1}{l}{FA} & \multicolumn{1}{l}{WA-L} & \multicolumn{1}{l}{\sys} \\ \hline
\multirow{2}{*}{UCF-101}    & \multirow{2}{*}{2.7B} & 100k      & 703s        & 389s      & 220s      \\
                            &                       & 200k      & 2757s       &     1272s      & 810s      \\ \hline
\multirow{2}{*}{Webvid} & \multirow{2}{*}{30B}  & 50k       &  112s           &    63s    &   39s        \\
                            &                       & 100k      &  396s           &   207s        & 125s          \\ \hline
\end{tabular}%
}
\caption{Inference latency comparison for 40 sampling steps using CFG~\cite{ho2022cfg}. 2.7B model runs on 1 H800 GPU and 30B model runs on 4 H800 GPUs with TP=4.}
\label{table:inference}
\end{minipage}
\vspace{-0.2in}
\end{table}

\noindent\textbf{User study.} We also conduct a user study with 30 volunteers, each blindly rating 15 sets of video samples generated by different methods in random order. Table~\ref{table:user_study} shows that \sys and FA generate videos with better quality to the human eye, significantly outperforming WA methods with WA-M receiving the lowest ratings.

\subsection{System Efficiency}
\label{sec:efficiency}
We evaluate \sys's efficiency compared to 3D FA and WA-L. We exclude WA-M in the comparison due to its consistently poor performance. 

\subsubsection{Training.} 

\begin{figure}[t]
  \centering
  \includegraphics[width=0.485\textwidth]{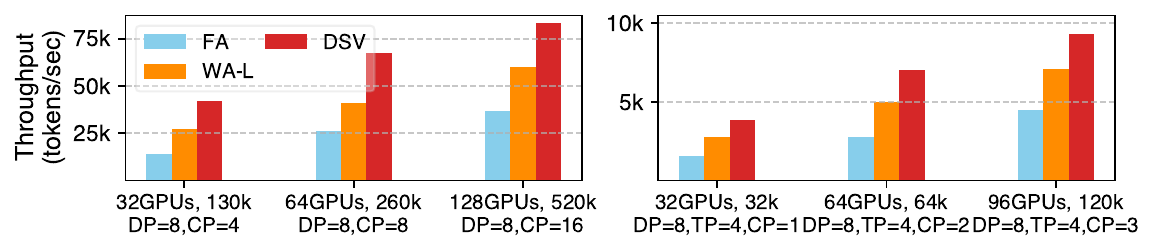} 
\caption{Training throughput comparison for 2.7B model with VideoGen (left) and 30B model with OpenVid (right). A tensor parallelism degree of 4 is used to accommodate the 30B model.}
  \label{fig:training_throughput} 
\end{figure}

In large-scale training, \sys significantly outperforms vanilla 3D FA and WA in training throughput. 
On the VideoGen dataset with a 2.7B model, \sys achieves 2.1-3.02$\times$ higher throughput over FA and 1.38-1.54$\times$ over WA-L on up to 128 GPUs while handling inputs of up to 520k tokens, as in Figure~\ref{fig:training_throughput}. 
On OpenVid with the 30B model, \sys outperforms FA by 2.06-2.53$\times$ and WA-L by 1.37-1.42$\times$.
FA suffers from high computational overhead due to its quadratic complexity in spatial and temporal dimensions. WA reduces this overhead by limiting attention to a fixed local window, achieving a fixed sparsity ratio ($\sim$70\% for WA-L) but fails to identify critical KV pairs. In contrast, \sys dynamically adapts to sparsity patterns across heads, achieving up to 98\% sparsity (see Figure~\ref{figure:videogen_sparsity} in Appendix~\ref{appendix:videogen_sparsity}). Additionally, \sys's optimized hybrid CP strategies addresses load imbalances and improves efficiency as well.

\subsubsection{Inference.} \sys also elevates inference efficiency via its low-rank sparsity predictors acquired during training and efficient kernel designs. 
This results in a 2.0-3.5$\times$ speedup over FA and a 1.4-2.6$\times$ speedup over WA-L, enabling faster inference without performance loss (see Table~\ref{table:inference}). Note CP is not used in the inference experiments.

\subsection{Deep Dive}

\subsubsection{\bf \em Critical KV prediction accuracy.} 
\begin{figure}[t]
    \centering
    \begin{minipage}[t]{0.25\textwidth}
       \centering
  \includegraphics[width=\textwidth]{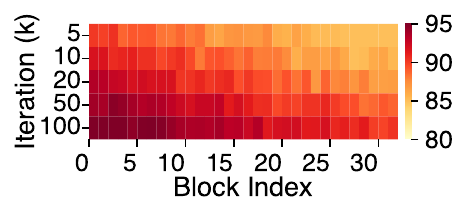}
\caption{Critical KV prediction accuracy.}
  \label{fig:prediction_acc} 
    \end{minipage}
    \hfill
    \hspace{-0.1in}
    \begin{minipage}[t]{0.215\textwidth}
      \centering
  \includegraphics[width=\textwidth]{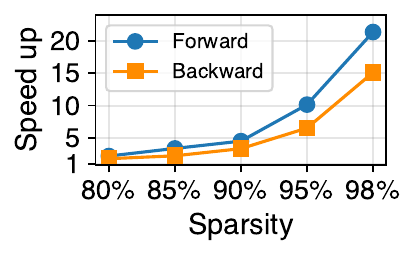} 
  \caption{The speedup over different sparsity of 2.7B model.}
  \label{fig:speed_up_different_sparsity} 
    \end{minipage}
   
\end{figure}

We evaluate our sparsity prediction for identifying critical KV pairs in a 2.7B VideoGen-trained model. As shown in Figure~\ref{fig:prediction_acc}, the prediction improves during the course of training and achieves over 90\% accuracy for most blocks after $\sim$100k iterations. 
The estimated KV pairs account for over 98\% of the total attention score compared to ground truth, as the most critical KV pairs are easily distinguishable. Although certain pairs are harder to identify, they seem to cause minimal effect on model performance.

\begin{figure}[t]
  \centering
  \includegraphics[width=0.45\textwidth]{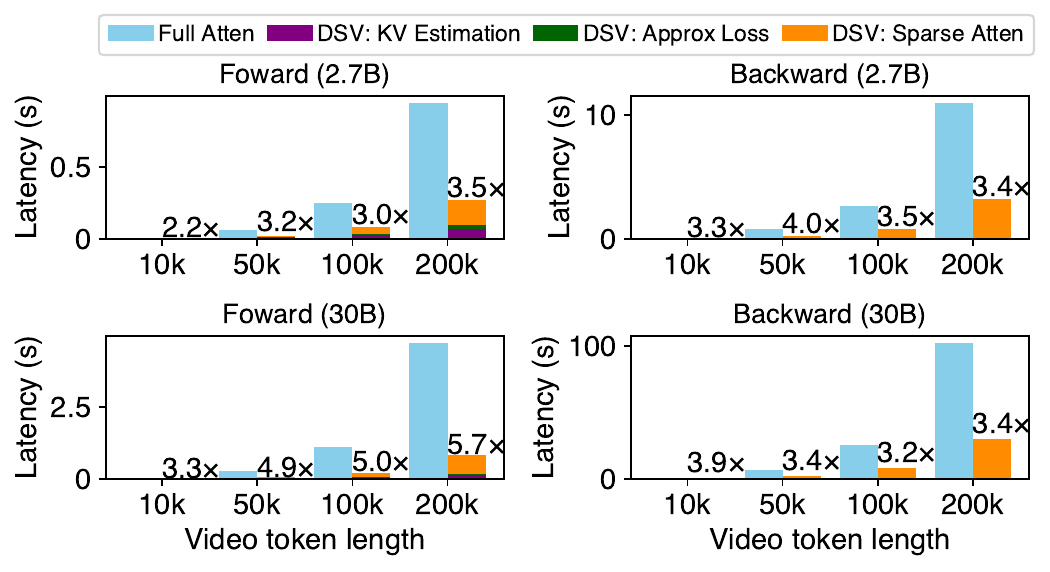} 
\caption{\sys's sparse attention module speedup and overhead breakdown with 90\% sparsity with different video token length. }
  \label{fig:kernel_speedup} 
\end{figure}

\subsubsection{\bf \em Kernel speedup and overhead breakdown.}
Figure~\ref{fig:kernel_speedup} shows \sys's efficient kernels deliver in total 2.2-5.7$\times$ and 3.3-4.0$\times$ speedups for forward and backward pass, respectively, at 90\% sparsity. This shows \sys's efficiency for long sequences.
Note the forward pass incurs overhead from estimating critical KV pairs and updating predictors.  
Figure~\ref{fig:speed_up_different_sparsity} further highlights \sys's speedup at varying sparsity levels for 100K sequences. It achieves over 15$\times$ speedup at 98\% sparsity compared to FA.

\begin{figure}[t]
  \centering
  \includegraphics[width=0.475\textwidth]{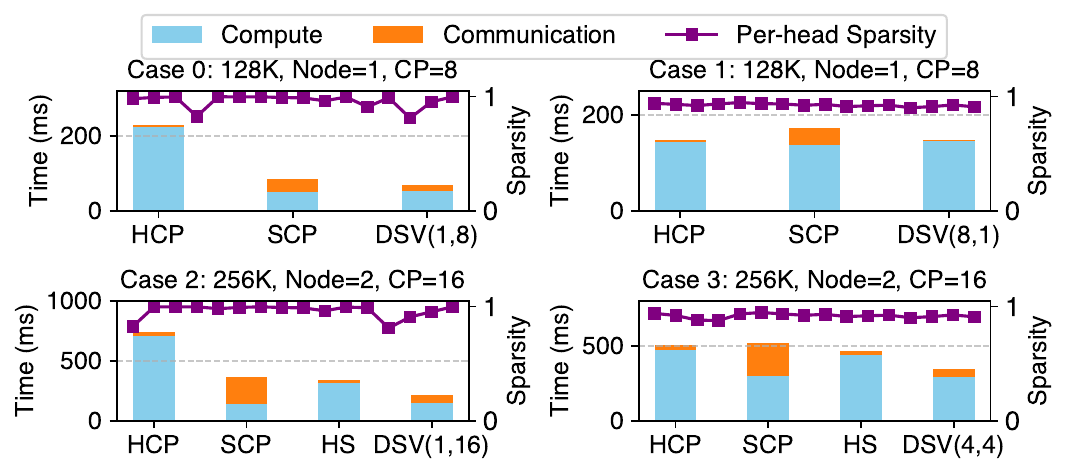} 
\caption{Comparison of context parallelism schemes for a 2.7B model, with timings based on the slowest GPU processing the attention operation. Line plots show head sparsity within each sampled block. ``HS'' combines standard intra-node HCP and inter-node SCP. \sys(x,y) denotes a sparse HCP degree of x and sparse SCP degree of y.}
  \label{figure:cp_ablation} 
\end{figure}

\subsubsection{\bf \em Hybrid sparse context parallelism} 
We compare \sys's hybrid sparse CP against conventional CP for the 2.7B model with varying sparsity patterns, GPU counts, and sequence lengths. 
We analyze the end-to-end execution time of a self-attention operation, focusing on the slowest GPU processing time within the CP group while accounting for sparsity disparities sampled from training. Figure~\ref{figure:cp_ablation} shows the results.
In cases 0 and 2, significant head sparsity outliers cause severe straggler effects in standard HCP and sparse HCP even after re-balancing. 
\sys opts to use SCP only in these cases to fully balance the workload and reduce the communication cost via selective KV gathering compared to standard SCP. 
In case 1, more evenly distributed sparsity allows sparse HCP to slightly outperform standard HCP while reducing communication overhead compared to SCP. In case 3, with moderately uneven sparsity, \sys optimally configures hybrid CP to balance computation, outperforming both HCP and HS, and lowering communication compared to seq-wise CP with its hybrid combination.

\subsubsection{\bf \em Threshold for critical KV}

\begin{figure}[]
  \centering
  \includegraphics[width=0.45\textwidth]{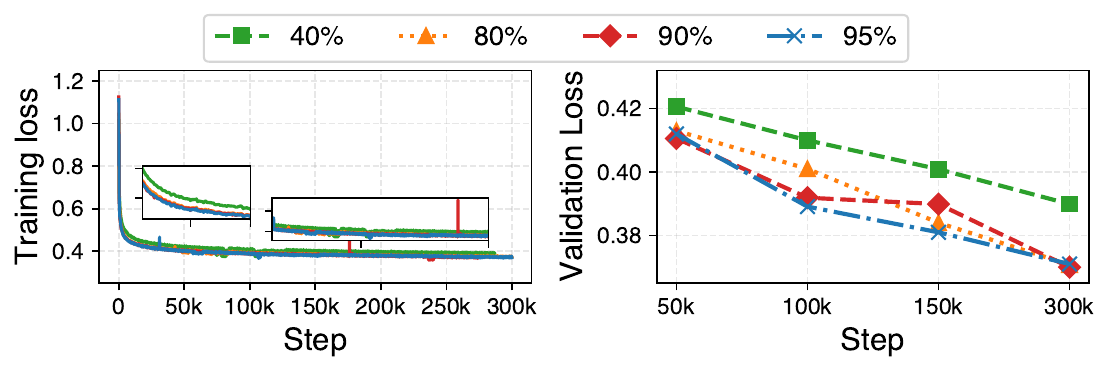} 
    \vspace{-0.1in}
\caption{The training and validation loss with different critical KV definition threshold on VideoGen. }
  \label{figure:critical_kv_threshold} 
\end{figure}

Recall we define critical KV with a fixed threshold, requiring them to contribute 90\% of the overall score (\Cref{sec:finding}). 
Here we analyze \sys's sensitivity to this threshold and find that it works for moderately large thresholds (>80\%).
A threshold set too low (40\%) omits too many critical KV and degrades model performance.
The results are shown in Figure~\ref{figure:critical_kv_threshold}.

\section{Limitations and Future Work}

\noindent\textbf{Finer-grained query-specific sparsity.} 
We currently use a uniform sparsity level for all queries in an attention head.
Exploring query-specific sparsity level may further improve the performance gain but also increase implementation complexity.

\noindent\textbf{Dynamic sparsity on pipeline parallelism.} Current video DiTs typically have under 30B parameters~\cite{hong2022cogvideo,yang2024cogvideox,polyak2024movie,OpenSora}, and tensor parallelism alone is sufficient. 
However, as model sizes grow, pipeline parallelism may become necessary, raising load balancing challenges due to dynamic sparsity across different stages. Efficient runtime management and block re-assignment strategies to handle these imbalances are an interesting problem.

\noindent\textbf{Kernel implementation.} Current Triton-based attention operation has slower backward performance than the CUDA version, as noted in our measurements and prior work~\cite{tritonflashattention,dao2023flashattention2}. Optimizing with CUDA could improve performance.

\section{Related Work}
\noindent\textbf{Video DiT training and inference.} Prior work explores novel attention mechanisms, such as 3D attention in expert transformers \cite{xu2024easyanimate}, hybrid spatio-temporal and full 3D attention \cite{HunyuanVideo}, and efficiency-focused designs with local attention and token merging~\cite{wang2024qihoot2xefficiencyfocuseddiffusiontransformer,pu2024efficient}. In contrast, \sys preserves model integrity by replacing only the attention module, ensuring compatibility with any architecture without structural changes. For inference, studies \cite{yuan2024ditfastattnattentioncompressiondiffusion,sun2024unveiling,kahatapitiya2024adaptive} have optimized computation by reusing cached activations via activation similarities across generation steps, which complements \sys's efficiency gains derived from its inherent sparsity.

\noindent\textbf{Attention sparsity exploration.} FlashAttention~\cite{dao2022flashattention1,dao2023flashattention2} provides block-level sparse attention but requires static sparsity patterns. Some work on LLMs has focused on sparsity during inference, particularly for long context. Methods like sliding window attention~\cite{mistral7b}, StreamingLLM~\cite{xiao2023streamllm}, and Minference~\cite{jiang2024minference} optimize efficiency using predefined or profile-based patterns. Besides, some studies explore trainable sparse attention~\cite{gao2024seerattention,yuan2025native}, using modules to learn block-level sparsity with training or fine-tuning. Similar techniques have been applied to video DiT inference~\cite{xi2025sparse,zhang2025fastvideotileatten}, though they typically impose fixed attention pattern and focus on inference in pre-trained models. 
In contrast, \sys is the first to adaptively exploit inherent sparse attention patterns for video DiT training while exploring sparse context parallelism.

\section{Conclusion}

We present \sys, a training acceleration framework leveraging video DiT attention sparsity. \sys employs a two-stage algorithm to enable adaptive sparse computation with specialized kernels. It also integrates hybrid sparsity-aware context parallelism to optimize sparse computation and communication across devices. These jointly yield up to 3.02$\times$ faster end-to-end training without sacrificing model quality on our 128-H800 testbed.

\newpage

\bibliographystyle{ACM-Reference-Format}
\bibliography{reference}

\clearpage
\twocolumn
\newpage

\appendix

\section{Hybrid sparsity-aware context parallelism algorithm}\label{appendix:algo}

The algorithm's pesudo code is listed in Algorithm~\ref{algo::cp}.
\begin{algorithm}[h]
\caption{Hybrid sparsity-aware context parallelism}\label{algo::cp}
\begin{algorithmic}[1]
\small
\Statex 

\Function{HybridSparseCP}{${Q}, {K}, {V}, {G_{hcp}}, G_{scp}$,}
\small
\Statex\quad\ {\textcolor{Maroon}{$\triangleright$ {Workflow from the perspective of each GPU (for simplicity, only the forward is shown).}}}
\Statex\quad\ {\textcolor{Maroon}{$\triangleright$ {$G_{hcp}$ and $G_{scp}$ are the sparse HCP and SCP communication groups this GPU belongs to. }}}

\If {$|G_{hcp}| >1$}
\State $Q,K,V$ $\gets$ \texttt{LoadBalanceHCP}($Q,K,V,G_{hcp}$)
\EndIf

\If {$|G_{scp}| >1$}
\State $K, V$ $\gets$ \texttt{SelectiveCommSCP}($Q, K, V, G_{scp}$)
\EndIf
\State $Output$ $\gets$ \texttt{SparseAttn}($Q,K,V$)

\Statex\quad\ {\textcolor{Maroon}{$\triangleright$ {Redistribute the output if any sparse HCP operations.}}}
\If {$|G_{hcp}| >1$}
\State $Output$ $\gets$ \texttt{UnevenAlltoAll}($output,G_{hcp}$)
\EndIf
\EndFunction

\Function{LoadBalanceHCP}{${Q}, {K}, {V}, Group$}
    \Statex\quad\ {\textcolor{Maroon}{$\triangleright$ {Get the balanced head-wise workload allocation for each GPU by solving the optimal allocation problem periodically.}}}
    
    \State $plan \gets \texttt{GetLoadBalancePlan}$(Group,Head\_Sparsity)
    \Statex\quad\ {\textcolor{Maroon}{$\triangleright$ {Permute $QKV$ along the head dimension to prepare the comm data layout.}}}
    \State $Q,K,V \gets \texttt{Permute\_and\_Split(Q,K,V,plan)}$

    \State $Q,K,V \gets \texttt{UnevenAlltoAll}(Q,K,V,Group)$
   \State \Return $Q,K,V$

\EndFunction

\Function{SelectiveCommSCP}{${Q}, {K}, V, Group$}
 \Statex\quad\ {\textcolor{Maroon}{$\triangleright$ {Get the global critical KV indices for each GPU with the sparsity predictor}}}
 
\State $Critical\_{KV}\_idx \gets \texttt{Estimate\_with\_Predictor}()$

 \Statex\quad\ {\textcolor{Maroon}{$\triangleright$ {Exchange information to decide which local KV pairs to send to other GPUs.}}}

\State $KV\_idx\_to\_send \gets 
\texttt{UnevenAlltoAll}(Critical\_{KV}\_idx)$
 \Statex\quad\ {\textcolor{Maroon}{$\triangleright$ {Prepare local KV to send to other GPUs}}}
\State $KV\_to\_send \gets \texttt{Prepare\_KV\_to\_send}(KV\_idx\_to\_send)$ 
\State $Gathered\_KV  \gets \texttt{UnevenAlltoAll}(KV\_to\_send)$
\State $K,V \gets \texttt{MergeKV}(K,V,Gathered\_KV)$
\State \Return $K,V$

\EndFunction

\end{algorithmic}
\end{algorithm}

\section{Implementation details}\label{appendix:imple}

\noindent\textbf{Asynchronous CPU offloading for large KV indices.} For long video tokens and relatively low sparsity ratios, sparse computation can generate large KV index tensors (e.g., approximately 2GB for a 50K video token length at 80\% sparsity). To mitigate GPU memory limitations, we employ asynchronous CPU offloading to transfer the KV index after attention computation and pre-fetch it during the backward pass of its next block, if necessary.

\noindent\textbf{Estimation kernel.} We perform matrix multiplication on CUDA cores rather than tensor cores due to its memory-bound 'slim' shape. Unlike standard tiling in matrix multiplication, each Streaming Multiprocessor (SM) conducts the multiplication for multiple full rows and selects the \topk indices online by Bitonic Select. However, large \topk sizes (e.g. 10K per query) can burden the SM's shared memory and reduce parallelism. To address this, we split the process into two stages: first, we perform multiplication and determine the \topk threshold for each query; second, we select indices based on the threshold.

\section{Sparsity change over \sys's training on VideoGen}\label{appendix:videogen_sparsity}

\begin{figure}[]
  \centering
  \includegraphics[width=0.43\textwidth]{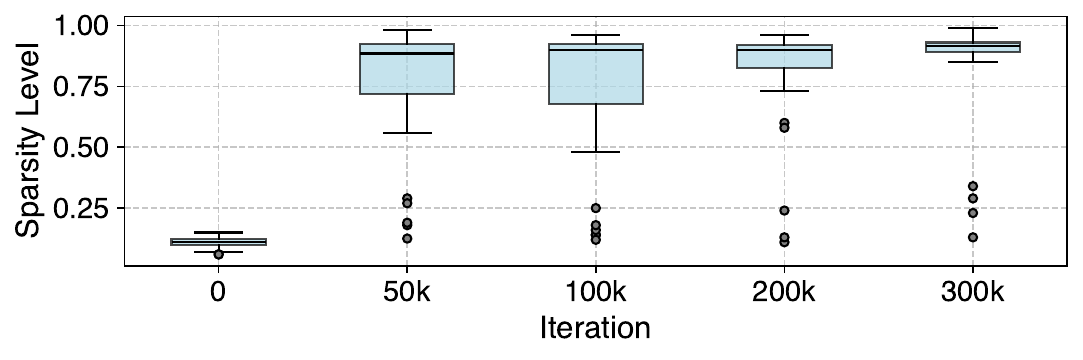} 
    \vspace{-0.1in}
\caption{The block-wise sparsity distribution throughout \sys training on VideoGen. }
  \label{figure:videogen_sparsity} 
\end{figure}

Figure~\ref{figure:videogen_sparsity} illustrates the evolution of block-wise sparsity throughout the training process of \sys on the VideoGen dataset. Notably, the low-sparsity outliers are mainly those from the initial and final blocks.

\end{document}